# CoCalc as a Learning Tool for Neural Network Simulation in the Special Course "Foundations of Mathematic Informatics"

Oksana Markova[1\[0000-0002-5236-6640\]], Serhiy Semerikov[2\[0000-0003-0789-0272\]] and Maiia Popel[3\[0000-0002-8087-962X\]]

[1] State Institution of Higher Education «Kryvyi Rih National University»
11, Vitali Matusevich st., Kryvyi Rih, 50027 Ukraine
`markova@mathinfo.ccjournals.eu`
[2] Kryvyi Rih State Pedagogical University
54, Gagarina Ave., Kryvyi Rih, 50086 Ukraine
`semerikov@gmail.com`
[3] Institute of Information Technologies and Learning Tools of NAES of Ukraine
9, M. Berlyns'koho St., Kyiv Ukraine
`popelmaya@gmail.com`

**Abstract.** The role of neural network modeling in the learning content of special course "Foundations of Mathematic Informatics" was discussed. The course was developed for the students of technical universities – future IT-specialists and directed to breaking the gap between theoretic computer science and it's applied applications: software, system and computing engineering. CoCalc was justified as a learning tool of mathematical informatics in general and neural network modeling in particular. The elements of technique of using CoCalc at studying topic "Neural network and pattern recognition" of the special course "Foundations of Mathematic Informatics" are shown. The program code was presented in a CofeeScript language, which implements the basic components of artificial neural network: neurons, synaptic connections, functions of activations (tangential, sigmoid, stepped) and their derivatives, methods of calculating the network`s weights, etc. The features of the Kolmogorov–Arnold representation theorem application were discussed for determination the architecture of multilayer neural networks. The implementation of the disjunctive logical element and approximation of an arbitrary function using a three-layer neural network were given as an examples. According to the simulation results, a conclusion was made as for the limits of the use of constructed networks, in which they retain their adequacy. The framework topics of individual research of the artificial neural networks is proposed.

# 1 Introduction

One of the necessary condition of fundamentalizing of computing education in higher educational and technical educational institutions is reorientation of basic information training from study rapid-changing technologies to a stable scientific basis of informatics. The leading role is played by computer modeling and numerical experiment [8], which simultaneously can be both methodological basis of informatics and learning methods of computing disciplines.

In the work [10] it is shown that effective way of fundamentalizing of informatic training students of pedagogical institutions is a *Mathematical Informatics.* Mathematical Informatics is the direction of scientific research. On the one hand, it is a component of theoretical computer science, where mathematical models and tools used for modeling and studying information processes in different spheres of human activity. On the other hand, it deals with the use of information systems and technologies for solving applied tasks. As an academic discipline *Mathematical Informatics* aims at mastering the basic models, methods and algorithms for solving problems arising in the field of intellectualization of information systems and considers the problem of the use of information, in particular mathematical models and information technologies for their research.

We have developed a special course "Foundations of Mathematical Informatics" which is intended for students of technical universities – future specialists in information technologies [9]. The content of the course is a combination of two interrelated components: theoretical and practical. The theoretical component is aimed to develop the students' ideas about data structures and algorithms that are the foundation of modern methodology of software development; methods for solving engineering and scientific tasks using numerical methods; the basic principles of coding and modulation of signals during the data transmission, signal processing, increase of noise immunity during data transfer via communication channels; basic methods of signal acquisition, decoding and detection errors by using various error-correcting codes; algorithmic aspects of number theory and their applications in modern cryptography. The practical aspect associated with the acquisition of skills to analyze, evaluate and select existing algorithms; to use methods and techniques of developing and evaluating the algorithms, develop new algorithms related to the design of hardware and software components of computer systems and networks; use existing and develop new mathematical methods for solving problems related to the design and using of computer systems and networks; to choose methods of computation that are resilient to errors; to solve linear and nonlinear algebraic equations and their systems; to apply interpolation and approximation; to make a selection of the method for integration of differential equations; to formulate and solve optimization problems; to apply the Kolmogorov–Arnold representation theorem to approximation arbitrary functions by three-layer neural network; to build the rings for the specified module; to apply the methods of error-correcting coding to data recovery when their injury; to build block ciphers; to build linear Bose-Chaudhuri-Hocquenghem codes; to build generating and testing polynomial for encoding and decoding cyclic codes; to apply Reed-Solomon's codes for data transmission in computer networks; to apply methods and tools to en-sure the security of programs and data in the design and operation of computer systems and networks; to consider the requirements of data protection; to create a software and hardware subsystem of cryptographic protection of data; to use the RSA algorithm and digital signatures for data transmission in computer networks; to create and manage by key information for the subsystems of the authentication; to use cloud technology for practical implementation of the basic methods of Mathematical Informatics.

There are 4 thematic modules in the content of the course.

In *the first substantive module* "Theory of algorithms" basic concepts and methods are discussed which are related to the analysis of algorithms (a machine with random memory access; analysis of the sorting algorithm by the inclusion; comparison of functions), algorithmic strategies (asymptotic analysis of upper and average complexity estimates of the algorithms; compare the best, average and worst estimates; O-, o-, ω0 and θ-notations; empirical measurements of the algorithm' efficiency; the overhead of algorithms by time and memory; recurrence relations and analysis of recursive algorithms; comparison of algorithms; the impact of the data structures and programming features on the algorithm efficiency; methods of algorithm development), algorithms design (value, classification and characteristics of sorting in the implementation of algorithms; simple sorting, their advantages and disadvantages; complex sorting and their advantages and disadvantages; comparison of simple and complex sorting).

In *the second substantive module* "Numerical methods" covers the basics of computer simulation (the concept of models and modeling; properties and classification of models; computer simulation features; statistical modeling features), tasks of linear and nonlinear algebra, approximation technique, methods of solution $1^{st}$-order ordinary differential equations; optimization technique (random search method, chord method, Golden section method; Fibonacci method; simplex search), neural networks and the task of pattern recognition (mathematical model of a neuron; the use of Kolmogorov–Arnold representation theorem to approximate arbitrary functions by three-layer neural network).

In *the third substantive module* "Coding theory" the mathematical foundations of coding theory, basic concepts of the error-correcting coding, linear codes, cyclic codes, Bose-Chaudhuri-Hocquenghem codes, Reed-Solomon codes, convolutional codes are discussed.

In *the fourth substantive module* "Basics of cryptography" the basic cryptographic system (symmetric and asymmetric) and their use for the management of cryptographic keys and digital signatures are discussed.

Special course final control of knowledge is a credit by the results of the current and the module control and presentation of individual education and research projects on the artificial neural networks building [2]. They was chosen due to the fact that, firstly, they are based on fundamental mathematical apparatus, and secondly, neural network modeling is one of the modern research directions in the field of mathematical informatics, and thirdly, the results which obtained during simulation can be applied in all substantive modules of the proposed special course.

## 2      The Aim and Objectives of the Study

Therefore, the aim of the study is to develop the individual components of the methodic of using cloud technologies as learning tool for neural network simulation in the special course "Foundations of Mathematic Informatics".

To accomplish the set goal, the following tasks had to be solved:

1. to justify the choice CoCalc as a learning tool of the foundations of mathematical Informatics for students of technical universities;
2. to develop demonstration models of artificial neural networks using various CoCalc components.

## 3      Literature Review and Problem Statement

One of the most powerful cloud technologies tools [1] is CoCalc (formerly known as SageMathCloud [6]) – a cloud based integrated version of the computer mathematics system Sage, hosted on Google's servers. CoCalc is not only the cloud based computer mathematics system, but also the system of support learning the mathematical and Computer Sciences subjects. The main components of CoCalc are:

1) Sage Worksheets – provides the ability to interactively run commands of Sage or programming (e.g., object-oriented and imperative) languages, such as C++ and HTML;

2) IPython notebooks (since 2016 – Jupyter Notebook) – timed session in Python programming language, the part of SciPy, scientific and engineering computing library. CoCalc provides the ability to multiple users to communicate through IPython notebooks in synchronous and asynchronous modes;

3) the workflow system in LaTeX with full support for sagetex, bibtex, etc.;

4) backup system – full save of all edited project files of the user every 2 minutes;

5) the replication system implies the preservation of each project in three physically separated data centers [7].

CoCalc provides opportunities of:

– interactive study of mathematics, natural and computer science;

– real time users collaboration;

– training: adding students, creating projects, monitoring of student's development, etc. using a cloud based educational materials;

– creating and editing of educational and academic texts using LaTeX, Markdown or HTML;

– adding your own files, data processing, presentation of results etc.

The presence of the 'Besides Sage Worksheets' tool in the composition of Jupyter Notebooks provides to the users of full access to classical Linux terminal [4].

The main CoCalc unit is a project. The user can create any number of independent projects of personal workspaces where the user stores resources of different types. The user can also invite others to collaborate in a joint project to provide open access to files or folders.

Each project is executed on the server CoCalc where it divides disk space, CPU and RAM with other projects. Free service plan provides using only those server resources that are free currently. In addition, when the user's project on the free service plan is not used for a few weeks, it is moved to secondary storage in order to free server resources and his re-starting will take significantly much more time than the user who paid for the service plan.

Project participants can combine their own computing and storage resources to improve the capabilities of the project as a whole and the reallocation of resources among themselves. To organize joint work with the resources of the CoCalc project is possible either at the level of individual resource, in particular of the worksheet, or project as a whole.

Opening of the share access at the level of individual resource is a web publication of the resource content in a read-only mode for all Internet users, which have link to this resource. The disadvantage of such publication is that the read-only user has no way to control the worksheet calculations, even if the author used the standard controls in it. However, if it necessary, the published worksheet can be copied or downloaded.

Organization of joint work at the level of the whole project is possible without/with the 'course' resource type. The first method involves connecting the participants to the project participants, who will have the ability to work together on existing educational resources of the project, or add new ones, invite other participants to communicate via text and video chats within the joint project. The contribution of each participant of the joint project in the solution of its tasks may be revised in the pages of history of the project or in the pages of his backups [3].

As a cloud subject-oriented environment, CoCalc in its composition contains both a computer mathematics systems and programming environments. The choice of a particular tool is carried out through binding to the file type or through the command of programming environment selection. At the stage of creating new files at project home directory, the user can choose the programming language. According to the choice, the environment is booting with internal compiler (interpreter).

The easiest way of handling CoCalc files is a Linux terminal mode. So, it is necessary to compile and run the developed program to test it. Files created as a result of program execution, become part of the student project in the CoCalc. Another method of executing programs in the CoCalc is directly on Sage worksheets. To do this, in the beginning of the cell, it necessary to specify one of the so-called "magic commands" (**%magic** below provides a full list of them). For example, **%coffeescript** executes the CoffeeScript code; the CoCalc is additionally define the printing function `print`. CoffeeScript code translates to JavaScript and runs directly in the browser, so the CoffeeScript program performance does not depend on the computing power of cloud servers.

## 4    Methodic of Using CoCalc as a Learning Tool for Neural Network Simulation

In the special course of the foundations of mathematical informatics using Coffee-Script can be considered such calculating-intensive tasks as creating and customizing of a neural network. Given the significant time required for this and the importance of the topic "Neural networks and pattern recognition" for the special course in general (such as topic which brings together computing and intelligent content lines), students are offered individual research task – development of an artificial neural network [5].

Artificial neural network is a mathematical model and also its software and hardware implementation, based on the principles of functioning of biological neural networks – networks of nervous cells of a living organism. This concept appeared in the study of processes that occur in the brain, and when we try to simulate these processes. After the development of the learning algorithms the models were used for practical purposes: in problems of prediction, pattern recognition, control problems, etc.

Artificial neural network is a system of interacted artificial neurons, interconnected through synapses. The input of the artificial neuron receives a set of signals, each of which is an output of another neuron. Each input is multiplied by weight coefficient of the synapse, all the components are summed, determining the activation level of a neuron as a scalar product of a vector input on the weight vector. The resulting value is measured by activation function, which normalizes the value in a given range: for polar activation function is [0; 1], bipolar [–1; +1].

Three-layer neural network is most commonly used; it architecture is presented in Fig. 1.

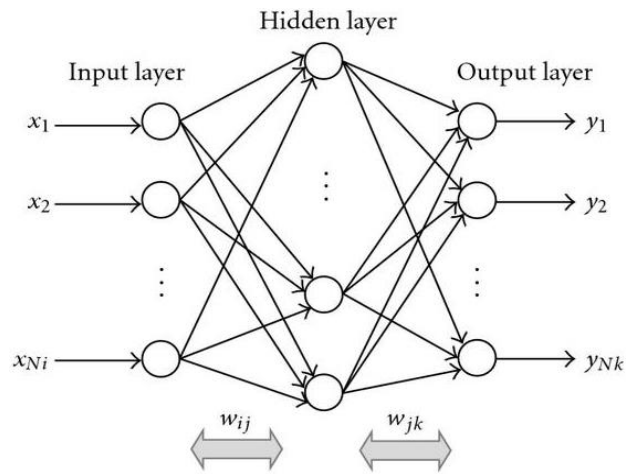

**Fig. 1.** The architecture of three-layer neural network

To develop a neural network, we offer students the following code in CoffeeScript:

```
%coffeescript
```

```
# Artificial neural network (based on Phillip Wang`s code
# https://github.com/lucidrains/coffee-neural-network)
class Synapse # synapse - connects two neurons
  constructor: (@source_neuron, @dest_neuron)->
    # initial weight is a random value within [-1;+1]
    @weight = @prev_weight = Math.random() * 2 - 1

class TanhGate # tangential activation function
  calculate: (activation)->
    math.tanh(activation)
  derivative: (output)-> # it's derive
    1 - output * output

class SigmoidGate # sigmoidal activation function
  calculate: (activation)->
    1.0 / (1.0 + Math.exp(-activation))
  derivative: (output)-> # it's derive
    output * (1 - output)

class ReluGate # Heaviside step activation function
  @LEAKY_CONSTANT = 0.01
  calculate: (activation)->
    if activation < 0 then activation *
      ReluGate.LEAKY_CONSTANT else activation
  derivative: (output)-> # it's derive
    if output > 0 then 1 else ReluGate.LEAKY_CONSTANT
```

The first of activation function is bipolar and corresponds to the hyperbolic tangent, the second is polar and corresponds to the logistic function. The latest activation function describes a polar stepped function.

```
class Neuron # artificial neuron
  # constants of learning rate and momentum
  @LEARNING_RATE = 0.1
  @MOMENTUM = 0.05

  constructor: (opts={})->
    gate_class = opts.gate_class || SigmoidGate
    @prev_threshold = @threshold = Math.random() * 2 - 1
    @synapses_in = []
    @synapses_out = []
    @dropped = false
    @output = 0.0
    @error = 0.0
    @gate = new gate_class()
```

```
dropout: ->
  @dropped = true
  @output = 0

calculate_output: -> # calculates the neuron response
  @dropped = false
  activation = 0
  for s in @synapses_in
    activation += s.weight * s.source_neuron.output
  activation -= @threshold
  @output =  @gate.calculate(activation)#

derivative: ->
  @gate.derivative @output

#calculation of weight coefficients of the output layer
output_train: (rate, target)->
  @error = (target - @output) * @derivative()
  @update_weights(rate)

#calculation of weight coefficients of the hidden layer
hidden_train: (rate)->
  @error = 0.0
  for synapse in @synapses_out
    @error +=
      synapse.prev_weight * synapse.dest_neuron.error
  @error *= @derivative()
  @update_weights(rate)

update_weights: (rate)->
  for synapse in @synapses_in
    temp_weight = synapse.weight
    synapse.weight += (rate * Neuron.LEARNING_RATE *
      @error * synapse.source_neuron.output) +
      (Neuron.MOMENTUM * ( synapse.weight -
      synapse.prev_weight))
    synapse.prev_weight = temp_weight
  temp_threshold = @threshold
  @threshold += (rate * Neuron.LEARNING_RATE * @error *
    -1) + (Neuron.MOMENTUM * (@threshold -
    @prev_threshold))
  @prev_threshold = temp_threshold
```

The network learning goal is to find the coefficients of neurons interconnections. During the learning process the neural network is able to identify complex dependencies between the input and output data, and perform generalization. It means that in case of successful learning the network will be able to return the correct result based on the data, which are absent in the training input, as well as incomplete and/or noisy, partly distorted data.

```
class NeuralNetwork # neural network
  @DROPOUT = 0.3

  # the constructor arguments is the type of the
  # activation function, the number of neurons on the
  # input layer, the number(s) of neurons on the hidden
  # layer(s), the number of neurons on the output layer
  constructor: (gate_class, input, hiddens..., output)->
    opts = {gate_class}
    @input_layer = (new Neuron(opts)
                    for i in [0...input])
    @hidden_layers = for hidden in hiddens
      (new Neuron(opts) for i in [0...hidden])
    @output_layer = (new Neuron(opts)
            for i in [0...output])
    for i in @input_layer
      for h in @hidden_layers[0]
        synapse = new Synapse(i, h)
        i.synapses_out.push synapse
        h.synapses_in.push synapse
    for layer, ind in @hidden_layers
      next_layer = if ind==(@hidden_layers.length-1)
        @output_layer
      else
        @hidden_layers[ind+1]
      for h in layer
        for o in next_layer
          synapse = new Synapse(h, o)
          h.synapses_out.push synapse
          o.synapses_in.push synapse

  train: (input, output)-> # neural network training
    @feed_forward(input)
    for neuron, ind in @output_layer
      neuron.output_train 0.5, output[ind]
    for layer in @hidden_layers by -1
      for neuron in layer
        neuron.hidden_train 0.5
```

```
  # feed the input signal through all network layers
  feed_forward: (input)->
    for n, ind in @input_layer
      n.output = input[ind]
    for layer in @hidden_layers
      for n in layer
        if Math.random() < NeuralNetwork.DROPOUT
          n.dropout()
        else
          n.calculate_output()
    for n in @output_layer
      n.calculate_output()

  # calculation result is on the output layer
  current_outputs: ->
    (n.output for n in @output_layer)

# cloning data
clone = (obj) ->
  return obj if obj is null or typeof (obj) isnt "object"
  temp = new obj.constructor()
  for key of obj
    temp[key] = clone(obj[key])
  temp
```

As an example, firstly we propose to examine a neural network for Boolean functions of two variables "OR". A feature of this example is that the input network served only two polar values 0 and 1, the output is also one of the two values.

Kolmogorov–Arnold representation theorem deals with persistence, under perturbation, of quasi-periodic motions in Hamiltonian dynamical systems.

In general, the phase space of a completely integrable Hamiltonian system of n degrees of freedom is foliated by invariant n-dimensional tori (possibly of different topology). This theory shows that, under suitable regularity and non-degeneracy assumptions, most (in measure theoretic sense) of such tori persist (slightly deformed) under small Hamiltonian perturbations. The union of persistent n-dimensional tori (Kolmogorov set) tend to fill the whole phase space as the strength of the perturbation is decreased.

Due to the Kolmogorov–Arnold representation theorem, in order for three-layer neural network reproduced any function of multiple variables, the dimension of the hidden layer should be at least more than 1 for twice dimension of the input. For this example, it is possible to reduce the dimension of the hidden layer from 5 to 2 due to the fact that there are only two possible values:

```
# Creating a three-layer neural network:
# 2 input neurons, 2 hidden and 1 output
```

```
nn = new NeuralNetwork(SigmoidGate, 2, 2, 1)

#training sequence consists of pairs "input – output"
pairs = [              #for example, for a disjunction
    [[0,0], [0]],
    [[0,1], [1]],
    [[1,0], [1]],
    [[1,1], [1]]
]

# Training limited by the number of iterations
numiter = 150000

for i in [0...numiter]
    err=0
    for pair in pairs
        nn.train pair[0], pair[1]
        nn.feed_forward pair[0]
        out=nn.current_outputs()
        for k in [0...pair[1].length]
            err+=(out[k]-pair[1][k])*(out[k]-pair[1][k])
    err=Math.sqrt(err/4)
    if i%1000==0
        print "Epoch ", i, ", error = ", err

# Network testing #1
for i in pairs
  nn.feed_forward i[0]
  print "Input ", i[0], ", calculated ",
       nn.current_outputs(), ", must be ", i[1]
```

The results of neural network testing is shown in Fig. 2.

```
Ітерація 140000 , помилка =  0.00421791796837043
Ітерація 141000 , помилка =  0.026060135533211935
Ітерація 142000 , помилка =  0.01662893256689042
Ітерація 143000 , помилка =  0.28026623651875215
Ітерація 144000 , помилка =  0.0010612101620747125
Ітерація 145000 , помилка =  0.2743559388471105
Ітерація 146000 , помилка =  0.028665800216671595
Ітерація 147000 , помилка =  0.027756609296225383
Ітерація 148000 , помилка =  0.0009493239126797038
Ітерація 149000 , помилка =  0.374065810007087847
Подаємо на вхід [0,0] , отримали [0.0010199996254682498] , повинно бути [0]
Подаємо на вхід [0,1] , отримали [0.9572457742977656] , повинно бути [1]
Подаємо на вхід [1,0] , отримали [0.9999319400884712] , повинно бути [1]
Подаємо на вхід [1,1] , отримали [0.466491093507080823] , повинно бути [1]
```

**Fig. 2.** Testing of neural networks for logic functions "OR" (in Ukrainian)

From Fig. 2 we can see that the number of iterations (150000) for network learning is too big, so that the computing process could be stopped while reducing the error to predetermined value beforehand. It is necessary to pay attention to the results of a calculation when the input network is supplied a pair (1; 1) – unlike the previous three tests, the obtained value is significantly differed from the needful. To resolve this error we offer to choose a different polar function of the activation – stepped.

The following example shows how to build neural network for random values of input and output:

```
# Creating a three-layer neural network:
# 3 input neurons, 7 hidden and 1 output
nn = new NeuralNetwork(SigmoidGate, 3, 7, 1)

# generate a new training sequence
data = []
count = 1000 #the number of pairs "input – output"

for i in [0...count]
    x1 = Math.random()*100-60
    x2 = Math.random()*100-40
    x3 = Math.random()*100-50
    data.push([[x1,x2,x3], [x1 + x2 - x3]])
```

The network architecture corresponds to the Kolmogorov-Arnold-Geht's-Nilsen's theorem, but the value at which it is proposed for training are not polar, as required by the logistic function. Bringing them to the desired range requires normalization, which is necessary to find limit values for input and output:

```
# find limit values for input and output
mininput = data[0][0][0]
maxinput = data[0][0][0]
minoutput = data[0][1][0]
maxoutput = data[0][1][0]

for i in [0...count]
    for j in [0...data[i][0].length]
        if data[i][0][j]<mininput
            mininput=data[i][0][j]
        if data[i][0][j]>maxinput
            maxinput=data[i][0][j]
    for j in [0...data[i][1].length]
        if data[i][1][j]<minoutput
            minoutput=data[i][1][j]
        if data[i][1][j]>maxoutput
            maxoutput=data[i][1][j]
```

Network training is performed on the normalized data (Fig. 3):

```
# training sequence, normalized in the range [0; 1]
normdata=clone(data)

for i in [0...count]
    for j in [0...data[i][0].length]
        normdata[i][0][j]=
           (data[i][0][j]-mininput)/(maxinput-mininput)
    for j in [0...data[i][1].length]
        normdata[i][1][j]=
           (data[i][1][j]-minoutput)/(maxoutput-minoutput)

numiter = 100000 # Training to complete iteration limit

for i in [0...numiter]
    err=0
    for pair in normdata
        nn.train pair[0], pair[1]
        nn.feed_forward pair[0]
        out=nn.current_outputs()
        for k in [0...pair[1].length]
            err+=(out[k]-pair[1][k])*(out[k]-pair[1][k])
    err=Math.sqrt(err/4)
    if i%1000==0
        print "Epoch ", i, ", error = ", err
```

```
Ітерація  96000 , помилка =  1.9951509319586445
Ітерація  97000 , помилка =  2.0141378667422836
Ітерація  98000 , помилка =  1.9155861398659084
Ітерація  99000 , помилка =  1.9405192614965148
Подаємо на вхід  [27.699390635066337,59.41341761361207,49.259104071956685] , отримали   36.35087315082319 , повинно
бути  37.85370417672172
Подаємо на вхід  [39.00633118239884,-16.21662884665037,-41.22289806481234] , отримали  -9.172790623682289 , повинно
бути  64.01260040056081
Подаємо на вхід  [27.747141959298915,-26.331914098465703,46.00194152219592] , отримали  -5.272461110228235 , повинно
бути  -44.58671366136271
Подаємо на вхід  [-23.11429081653037,-24.202363779809698,-0.16588587495485285] , отримали  -30.680435299350293 ,
повинно бути  -47.15076872138522
Подаємо на вхід  [-11.79389853611631,23.72835228983797,-19.9927199117117] , отримали  -3.1010809642714605 , повинно
бути  31.92717366543336
Подаємо на вхід  [27.519457157784757,-0.8885653362011041,35.02090493243577] , отримали   37.646162889115914 , повинно
бути  -8.3900131108522117
Подаємо на вхід  [5.519824002791353,25.860920751786693,35.38865410980026] , отримали   21.525788294005025 , повинно
бути  -4.0079093552222105
Подаємо на вхід  [-59.39481245405906,31.465997970337654,46.95086766798531] , отримали  -26.363806615506718 , повинно
бути  -74.8796821517067
Подаємо на вхід  [16.04025015123746,-3.9688678307271914,3.623055486515561] , отримали   15.30852052831304 , повинно
бути  8.448326833994706
Подаємо на вхід  [11.559482975934174,38.75477763182535,-41.237508936767384] , отримали   69.23303789586686 , повинно
бути  91.5517695445269
```

**Fig. 3.** The results of testing the neural network for adding function (in Ukrainian)

While testing a network, perform the reverse process of denormalization:

```
# Network testing #2
for i in [0...count]
  nn.feed_forward normdata[i][0]
  res=nn.current_outputs()
  print "Input ", data[i][0], ", calculated ",
    res[0]*(maxoutput-minoutput)+minoutput,
    ", must be ", data[i][1][0]
```

In Fig. 3 shows the test results.

While discussing the test results, it is advisable again to pay attention to the values that differ significantly from the etalons. We first recall that the input values generated randomly in the following ranges: $x_1 \in [-60; 40)$, $x_2 \in [-40; 60)$, $x_3 \in [-50; 50)$. Analysis of the results shows that, then closer the input values to the range limits, then greater the difference of the result from the etalons. This provides an opportunity to do conclusion on the boundaries of application of the constructed network in which it maintains adequacy.

## 5     Conclusions

1. The special course "Foundations of Mathematic Informatics" for students of technical universities – future IT-experts aimed and directed to breaking the gap between theoretic computer science and it's applied applications: software, system and computing engineering. In this regard, their fundamental foundations are implemented using modern programming languages and cloud technologies tools.

2. One of the leading cloud technology learning tools of the special course is CoCalc – the mathematical software system that provides the ability to support all sections of the special course in an unified mobile mathematical environment. Despite the fact that Python is the most often used programming language in CoCalc, a program realization examples represented by the extension of browser-based JavaScript language, which provides to developed software a higher level of mobility.
3. The central theme of the special course is the neural network simulation – a traditional technique for modeling natural neural networks, which had a significant impact on all stages of the development of computer and software engineering. The most versatile neural networks architectures and their application to the problems of modeling the basic logical elements of the computer system and identifying hidden dependencies are discussed in paper.
4. The final evaluation for the special course includes the presentation of individual education and research projects on the artificial neural networks building. The framework project's topics involves modeling continuous, discrete-continuous and discrete neural networks for solving problems of circuit synthesis, time series forecasting, pattern recognition, functions approximation, dependency identification, medical diagnostics, decision-making under conditions of incomplete data, data compression, unknown data restoration, clustering, automated control etc.